\pdfoutput=1
\documentclass[modern]{aastex61}

\received{\today}
\revised{\today}
\accepted{\today}

\submitjournal{ApJ}

\newcommand{\beq}{\begin{equation}}
\newcommand{\eeq}{\end{equation}}
\newcommand{\Mdot}{\dot{M}~}
\newcommand{\hii}{H~{\sc II}~}
\newcommand{\kms}{\mbox{ km s$^{-1}$}~}

\newcommand{\Mo}{\mbox{M$_{\odot}$}~}
\newcommand{\Moy}{\mbox{M$_{\odot}$ yr$^{-1}$}~}

\newcommand{\Ro}{\mbox{R$_{\odot}$}}

\shorttitle{Common Envelope Shaping of Planetary Nebulae}

\shortauthors{Garc\'{\i}a-Segura et al.}

\begin{document}

\title{Common Envelope Shaping of Planetary Nebulae}

\correspondingauthor{Guillermo Garc\'{\i}a-Segura}
\email{ggs@astrosen.unam.mx}

\author{Guillermo Garc\'{\i}a-Segura}
\affiliation{Instituto de Astronom\'{\i}a, Universidad Nacional Aut\'onoma
de M\'exico, Km. 107 Carr. Tijuana-Ensenada, 22860, Ensenada, B. C., Mexico}

\author{Paul M. Ricker}
\affiliation{Department of Astronomy, University of Illinois, 1002 W. Green St., Urbana, IL 61801, USA}

\author{Ronald E. Taam}
\affiliation{Institute of Astronomy and Astrophysics, Academia Sinica, Taipei, 10617, Taiwan}
\affiliation{Department of Physics and Astronomy, Northwestern University, 2145 Sheridan Road, Evanston, IL 60208, USA }

\begin{abstract}

The morphology of planetary nebulae emerging from the common envelope phase of binary star evolution is investigated.
Using initial conditions based on the numerical results of hydrodynamical simulations of the common envelope phase it is found that the shapes and sizes of the resulting nebula are  very sensitive to the  effective temperature of the remnant core, the mass-loss rate at the onset of the common envelope phase, and the mass ratio of the binary system. These parameters are related to the efficiency of the mass ejection after the spiral-in phase, the stellar evolutionary phase (i.e., RG, AGB or TP-AGB), and the degree of departure from spherical symmetry in the stellar wind mass loss process itself respectively.  It is found that 
the shapes are mostly bipolar in the early phase of evolution, but can quickly  transition to elliptical and barrel-type shapes. Solutions for nested lobes are found where the outer lobes are usually bipolar and the 
inner lobes are elliptical, bipolar or barrel-type, a result due to the flow of the photo-evaporated gas from the equatorial region. It is found that the lobes can be produced without the need for two distinct mass ejection events. In all the computations, the bulk of the mass is concentrated in the orbital or equatorial plane, in the form of a large toroid, which can be either neutral (early phases) or photoionized (late phases), depending of the evolutionary state of the system. 
\end{abstract}

\keywords{ Stars: Evolution ---Stars: Rotation ---Stars: RG, AGB and 
post-AGB ---(Stars:) binaries: general  ---(ISM:) planetary nebulae: general}

\section{Introduction} \label{sec:intro}

Investigations of the formation of planetary nebulae (PNe) have significantly progressed since the pioneering work by Kwok, Purton \& Fitzgerald (1978). In particular, several explanations have been advanced for producing the shapes of PNe, with a specific emphasis on the formation of bipolar structures. In some of the early models, the presence of a dense medium in the equatorial latitude is invoked, which collimates a fast stellar wind that is emitted in the late stages when the remnant core is revealed (Calvet \& Peimbert 1983; Balick 1987; Icke 1988; Icke et al. 1989; Mellema et al. 1991; Frank \& Mellema 1994). Subsequent studies focused on the manner in which a dense medium is formed in the equatorial region
of the asymptotic giant branch (AGB) star.  Suggestions include, for example, the geometry of the stellar wind is affected by stellar rotation (Ignace et al. 1996; Garc\'{\i}a-Segura 1997; Garc\'{\i}a-Segura et al. 1999), and magnetic forces (Matt et al. 2000). In this context, Garc\'{\i}a-Segura et al. (2014) showed that the 
bipolar PNe formation cannot result from the evolution of a single AGB star, since most of the spin angular momentum is lost by the stellar wind at the thermal pulsing, asymptotic giant branch (TP-AGB) phase.

The need for a binary companion which provides a reservoir of angular momentum has also been
the subject of a long standing debate during these past four decades, since an asphericity of the AGB wind could naturally arise from the gravitational interaction of AGB star with a companion star in several ways. One hypothesis invokes spin-up of the envelope of the AGB star by tidal forces (Livio 1994; Soker 1998). In this context, Garc\'{\i}a-Segura et al. (2016) demonstrated that bipolar PNe formation cannot result from the spin-up of the AGB star alone, since, the gained angular momentum is lost by the stellar wind at the TP-AGB phase. The above study suggests that it would be more natural to produce a bipolar PN via a common envelope evolution scenario (Livio 1993; Soker 1997; Sandquist et al. 1998;
Nordhaus \& Blackman 2006; de Marco 2009; Ricker \& Taam 2012; de Marco et al. 2013).

In this paper, we report on the results of two dimensional hydrodynamical simulations of the formation of PNe which, for the first time, consider an evolutionary state in which the system has undergone a common envelope phase. 
In the next section, the numerical scheme and
physical approximations as well as the inputs in our calculations are described.  The results of the numerical simulations for a post-red giant (post-RG) nebula are presented in \S~3 and for  post-AGB nebulae for two different initial masses in \S~4.  The influence of the mass ratio of the binary and of thermal pulses is described in \S~5 and \S~6 respectively. In \S~7 the effect of the initial effective temperature of the central star after envelope ejection on the planetary nebula morphology is described, and our models are qualitatively compared with observed PNe in \S~8. Finally, we discuss the numerical results in \S~9 and provide the main conclusions in the last section.

\section{Numerical methods and physical assumptions}

The simulations have been performed using the magneto-hydrodynamic code
ZEUS-3D (version 3.4), developed by M. L. Norman and the Laboratory for
Computational Astrophysics. It is a finite-difference, fully explicit,
Eulerian code descended from the code described in Stone \& Norman (1992) and
in Clarke (1996). A simple approximation is used to derive the location of the ionization
front in this study (Garc\'{\i}a-Segura et al. 1999).  
This is done by assuming that ionization equilibrium applies at all times, 
and that the gas is fully ionized inside the \hii region.
The models include the Raymond \& Smith (1977) cooling curve above $10^4$ K.
For temperatures below $10^4$ K, the shocked gas region is allowed to cool based on the radiative cooling curves given by Dalgarno \& McCray (1972) and MacDonald \& Bailey (1981). Finally, the photoionized gas is always kept at $10^4$ K,
so no cooling is applied to the photoionized regions (unless a shock is present within the photoionized region). The minimum temperature allowed in all models is set to $10^2$ K. 

The computational grid consists of $200 \times 200$ equidistant zones in $r$ and $\theta$ respectively, with an angular extent of $90^{\circ}$, and
initial radial extent of $1.9 \times 10^{13}$ cm (= 1.266 AU). A self-expanding grid technique (Garc\'{\i}a-Segura et al. 2005) has been employed in order to allow the spatial coordinates to grow by several orders of 
magnitude.  A shock-tracking routine evaluates the maximum expansion velocity for each 
forward shock ($v_s$) near the edge of the grid  and produces a self-expanding grid moving at a velocity $v_g(i) = v_s [r(i)/r_s] $, where $v_g(i)$ and $r(i)$ are the velocity and position of
each grid zone in the r-coordinate and $r_s$ is the position of the shock wave.
Thus, the final grid size depends on the dynamical evolution of each individual run.

As input conditions we adopt the numerical data describing the outcome of the common envelope phase from the computations presented in Ricker \& Taam (2012). This computation used the AMR code FLASH 2.4 (Fryxell et al. 2000).
The input data are originally three dimensional in Cartesian coordinates. However, 
computing the nebular evolution with a similar scheme up to 10,000 years is prohibitive for actual supercomputers, since 56.7 days of the common envelope 
evolution calculation took the equivalent of 512 cores for 792,000 core hours on the 
Ranger cluster at the Texas Advanced Computing Center. Hence, the computations presented in this paper are performed in two spatial dimensions ($r,\theta$). The densities and temperatures  are averaged over $\phi$, and the 
velocity components ($v_r, v_{\theta}, v_{\phi}$) are averaged over $\phi$ using density weighting. 
The averages are done using random sampling.
Figure 1 illustrates the averaged density distribution at 56.7 days of the calculation, which shows the ejected envelope and is the initial condition for this study.

We have implemented a correction to the dynamics at early phases, since
the numerical data adopted from Ricker \& Taam (2012), when used as initial conditions in our code, produces a fast flow of mass from
the equator to the polar region  due to the large pressure gradient. This behavior results from the fact that ZEUS-3D does not include self-gravity in spherical coordinates, and the gravitational force associated with the toroidal distribution of gas at the equator 
(the ejected envelope) is not included. 
To circumvent this lack of consistency between the two codes, we initially set the force terms in the momentum equation equal 
to zero, and allow a smooth transition to a full dynamics in a period of time (40 years).  
In this way, there is no tendency for the pressure gradient to cause matter to 
flow to the pole.  That is, it is implicitly 
assumed that the gravitational force due to the core and the self gravity of the 
common envelope balances the pressure force.  This leads to a determination for the 
temperature distribution of the ejected envelope as a function of time and space.  

The inflow of mass into the computational grid representing the stellar wind from the remnant core is treated as an inner boundary condition,  at the first two innermost
radial zones, and is taken from Villaver et al. (2002b), based on the stellar
evolution models of post-AGB stars (1 \Mo and 2.5 \Mo) from Vassiliadis \& Wood (1994).
For the models R1 and R2 (RG cases), the mass-loss rates and  photoionizing photons
have been decreased according to the total luminosity (a factor $\sim  0.03$ lower),
since the RG core is less massive and less luminous than the AGB core.
For the outer, expanding  boundary, we use mass-loss rates for the RG and AGB  
winds from Villaver et al. (2002a), which are based on Reimers (1975) and 
Vassiliadis \& Wood (1993). External to the RG and AGB winds, the interstellar medium density is set to 0.01 $cm^{-3}$, which corresponds to densities external to the Galactic spiral arms (Villaver et al. 2012). 

The influence of the binary companion can produce a density enhancement in the equatorial region, either by gravitational focusing (Gawryszczak et al. 2002), Wind-Roche-Lobe overflow (WRLOF) (Podsiadlowski \& Mohamed 2007; Mohamed \& Podsiadlowski 2012), or Roche-Lobe overflow 
(RLOF) (Bermudez et al., in preparation). However, none of these studies provide 
an analytical description to prescribe the density and velocity at our boundary. As a consequence, we have chosen to adopt a correction for the wind-compressed zone according to Bjorkman \& Casinnelli (1993)  and Garc\'{\i}a-Segura et al. (1999). This correction produces an equatorial density enhancement on the RG and AGB wind which is similar to the one produced by a binary.  We use a critical rotation of 97\%  for this purpose. Future computations will include a more realistic prescription, since density gradients from the pole to equator can reach values of three orders of magnitude (model A, for a mass ratio of 0.6,  in Gawryszczak et al. 2002). 

Ricker \& Taam (2012) computed the case of a binary initially consisting of a 1.05 \Mo  RG star containing a 0.36 \Mo degenerate core
with a 0.6 \Mo companion. No other case was computed at such a resolution. 
Since we are interested in PNe descending from AGB stars, 
for simplicity, we assume that the ejected gas is the same for all stellar masses.
However, the velocity components for the ejected envelope are reduced in the AGB cases by half, to account for a smaller
escape velocity. The gravity produced by the two
stellar cores is included in the calculation as a central, point mass. 

The simulations are run in a single processor mode on a Supermicro server. 
At the resolution of $200\times 200$ , they take from ${\sim 1}$ to ${\sim 3}$ days, depending of the initial conditions. These are quite long computational times for the used resolution and result from the fact that the initial grid is of very  small physical dimensions ($\sim 10^{13}$ cm) and the physical evolution times
are very long ($10^4$ yr). For a grid  resolution of $400 \times 400$, the computations require from ${\sim 8}$ to ${\sim 24}$ days, and a $1000 \times 1000$ will require from ${\sim 125}$ to ${\sim 375}$ days. Although we are not physically  limited by the resolution used, it is not useful to run many models in the highest
resolution because of the cost in time. For this reason, we only calculate two models in a $400 \times 400$ resolution, and show that the used resolution is sufficient in our qualitative study. 

\section{Nebula formation and evolution: post-RG nebulae}

The simulations presented in this study start ($t=0$) when the RG envelope has
been partially ejected, i.e.,  after 56.7 days of the common envelope evolution in the 
calculation of Ricker \& Taam (2012).  We note that only $\sim$ 26\% of the envelope mass is unbound, corresponding to the outer, filamentary wings visible in Figure 1, which are ejected at $\sim$ 50 \kms. The rest of the envelope mass, either contracts or remains in a nearly keplerian or sub-keplerian orbit. This mass is the densest, inner part of the 
wings visible in Figure 1.  The ejected envelope is characterized by a high  temperature and averages $\sim 10^5$ K.

As the evolution proceeds the bound envelope contracts back to the stellar interior or collapses into a circumbinary disk, and the unbound part forms an expanding toroidal structure surrounded
by a bow shock. This is already visible in frame A of Figure 2. Meanwhile, the inner fast wind is already developed at 140 days. During this initial phase, the cooling of the envelope is mainly adiabatic, since the gas at high densities remains optically thick. The remnant core is assumed to have an initial effective temperature of 29,000 K. Although very uncertain, we have chosen this value in order that a moderate ($\sim 600$ \kms) initial wind velocity is produced that is greater than
the escape velocity from the central binary system ($\sim 500$ \kms) at the inner
boundary. The adopted mass-loss rate of $10^{-9}$ \Moy for the RG star is based on the Reimers (1975) prescription.

After 1000 days of evolution (frame B of Figure 2), a well established bipolar
nebula is produced,  
and the ejected torus is converted into an excretion disk due to angular momentum conservation. The gas that is ejected with a large $v_{\phi}$ component within a certain angle from the equator, has an orbital plane which intersects the center of mass and is tilted by the same amount with respect to the equatorial plane. Thus, soon thereafter, the gas crosses the equatorial plane and encounters gas  ejected from the opposite hemisphere, accumulating
at the equator, in a manner similar to the wind compressed model described in 
Bjorkman \& Casinelli (1993).

Once the excretion disk is formed, it contains most of the ejected mass and will play an important role in the nebular dynamics. This is evident 
already at 100 years of the evolution (frame C in Figure 2), where the
photoionization produces an important photoevaporated wind from the excretion
disk which opposes the inner fast wind, piling up the gas and  producing a nested, inner nebula. In the remaining frames of Fig. 2, it can been seen that the effect of photoevaporation  is quite dominant on the evolution of the nebula. 

We note that the external bipolar lobes are formed by swept-up gas from the RG wind,
which is the classical mechanism for forming lobes in the interacting wind scenario (Kwok et al. 1978), but the formation of the inner lobes is completely different and is produced directly by the envelope gas. Moreover, the gas
that forms the inner lobes originates from the deepest regions of the ejected
envelope, which corresponds to the inner part of the excretion disk.

The long term evolution of the nebula (frames G and  H of Figure 2) is dictated by the expansion  of the remnant excretion disk. At this point, the nebula expands into a flat density ISM. 

\section{Nebula formation and evolution:  post-AGB nebulae}

To examine the differences associated with the evolutionary state of the giant star, we follow the evolution of a post-AGB nebula, 
assuming that the distribution of the ejected envelope is similar to the one for the RG case. This is plausible since the morphology of 
the outflowing matter from the
simulation by Ricker \& Taam (2012) is generic to outflows resulting from the 
common envelope phases (Taam \& Sandquist 2000).

Both models in Figures 3 and 4 (A1 and A2)  have similar initial conditions
to model R2 (Figure 2), except for the reduction of the initial velocities 
(by 50\%) to account for smaller AGB escape velocities. 

The difference between A1 and A2 is the different stellar mass cores,  0.569 and 0.677 \Mo respectively, which in turn produce more 
energetic winds for case A2, but also of shorter duration (see Figure 3 in Vassiliadis \& Wood 1994 to visualize
the HR diagram, and see Figures 3 and 4 in Villaver et al. 2002b for the
wind momentum, kinetic energy and ionizing photons).
These differences in momentum and energy input result in a factor $\sim$ 2 larger nebula size in model A2 than in model A1 during the evolution. 

Both models produce bipolar shapes in the outer lobes, with larger sizes
and elongated morphology in the case of model A2. For an assumed initial effective temperature of 29,000 K of the remnant core, model 
A1 ends when $T_{\rm eff} \sim 67,000 $ K, 
and A2 when $T_{\rm eff} \sim 115,000 $ K.
The final wind velocity for model A1 is $\sim 2,300 $ \kms and   
$\sim 14,000 $ \kms for model A2. We note that the models assume small mass-loss rates ($10^{-8}$ \Moy) for the AGB phase, i.e., the common envelope events take place prior to the thermal pulsing phase. This
corresponds to 50\%  and 85\% of the AGB lifetime for models A1 and A2 respectively. We will study the TP-AGB case below, where 
the AGB radius can exceed 400 \Ro \, in the case of model A2 (see Figure 8 in 
Garc\'{\i}a-Segura et al. 2014).

A number of different morphologies are found in the inner, nested lobes, 
including bipolar (frame C in Figure 3), elliptical (frame H in Figure 3) and barrel-type (frame D in Figure 4). 
Frames A and B in Figures 3 and 4 reveal 
non-linear, thin shell instabilities (Vishniac 1994) because of a nearly 
isothermal, fully radiative dynamics. 
These instabilities appear when both, the outer and the inner shocks are radiative, and, in the case of the inner shock, the fast wind impacts directly onto the swept-up shell.
We note that radiative shocks occur close to the star because of the high densities
(radiative cooling is proportional to $n^2$)
which scale as $r^{-2}$, happening in the early phases of PNe. 

\section{Studies on the binary mass ratio}

The circumbinary density distribution resulting from the common envelope evolutionary phase is very promising 
for producing bipolar PNe, but the previous phase just prior to this event is very likely 
responsible for  determining the bipolarity in the outer lobes. We have obtained evidence in support of this 
interpretation by comparing the influence that the binary mass ratio has on the resulting morphology.
Although we do not directly compute the influence of the gravitational effect from the secondary star on
the previous wind, we introduce an equatorial density enhancement that mimics this effect. Consequently, 
a large mass ratio produces a large density enhancement, while a very small mass ratio produce a negligible effect, i.e., a spherical wind. 

Figure 5 shows two models, R1 and R2, with identical initial conditions and identical 
stellar cores (i.e., same winds and ionizing fluxes), but a difference in the treatment of the external, 
preexisting RG wind. In R1, we assume 
a spherical RG wind, while in R2, we introduced an equatorial density enhancement based on the analytical equations in Garc\'{\i}a-Segura et al. (1999), using 97\% of critical rotation as an example. 

The resulting form of the outer lobes completely differs in the two models. While the outer
lobes in R1 are rather spherical in shape at 100 years, the outer lobes of R2 are characterized by a classical bipolar shape.
Although  the fast wind is collimated to some degree by the excretion disk, as can be seen in the left frames
of Fig. 5 at 1000 days, this collimation does not result in a bipolar nebula. The resulting bipolar morphology of the outer lobes is attributed to the fact that the expansion of the nebula is favored in the polar direction due to the density gradient.
This result provides an excellent diagnostic probe to study the type of binary star
precursors involved in producing aspherical PNe.
We note, however, that the shape of the inner lobes do not provide such a probe. Here, the inner lobes of R1 have a 
bipolar shape in the form of a waist which resembles the morphology of Abell 63 (Mitchell et al. 2007), but the outer lobes are nearly spherical. Hence, the description of the shape of the outer lobes can, in principle, be used to provide diagnostic information of the precursor binary system. We note that the case of model R2 appears similar to Abell 65, which exhibits an external bipolar shape in the outer lobes,
but also an elliptical shape in the inner lobes (Huckvale et al. 2013).

\section{Studies on the thermal pulses}

The description for the onset of the common envelope evolutionary phase depends on the orbital separation of the binary system. For the largest possible separations in which two stars can interact, 
common envelope evolution can occur when the radii of stars are near their maximum, occurring only at the TP-AGB phases for the intermediate mass stars leading to the formation of PNe. During these 
stellar evolutionary phases, the mass-loss rates from AGB stars are characterized by periodic and 
dramatic increases, and such conditions produce large changes in the circumstellar density environment 
into which PNe expand (see Villaver et al. 2002a, 2002b). As a consequence, the common envelope 
evolutionary stage is not reached in the same manner during the initial phases of the TP-AGB,
during periods of maximum 
mass-loss rates (pulses), or when the star is in a quiescence state in between pulses. Hence, 
the initial conditions describing the density distribution, which determines the PN formation 
after a common envelope evolution is rather different in the thermally pulsing state. 

Motivated by these aspects of the mass loss process during this evolutionary phase, we have performed several additional simulations to determine the main differences on the PN structure due to the magnitude of the mass-loss rates prior to the common envelope phase. In 
particular, we have considered
three different values for the AGB mass-loss rates, $10^{-6}$, $10^{-7}$ and $10^{-8}$ \Moy.
The above values mimic the mass-loss rates during pulses, quiescence states and prior to thermal pulses respectively.

The models A4, A5 and A6 are shown in Figure 6, in decreasing order of the mass-loss rate. For 
this study, a central star with initial $T_{\rm eff}$= 55,000 K is used, in order to drive the 
expansion of the wind for the conditions for model A4. The initial velocity of the fast wind velocity is taken to be 1680 \kms. 

The numerical results show that  the larger the AGB mass-loss rate,
the smaller will be the 
external lobes. This is quite revealing since one would expect that stellar evolution 
theory would predict the same size of nebula once the progenitor mass is fixed. This is likely to be the case for isolated stars, but is completely changed when a common evolution phase occurs, in which the size of the nebula
will depend on the synchronization of the onset of the common envelope phase with the thermal pulse.  

The internal lobe structure is also affected. Specifically, the higher the AGB mass-loss rate, the larger is the density of the AGB wind into which the common envelope ejecta collides.  This leads to a greater deceleration in the expansion of the excretion disk. 
Thus, for a high mass-loss rate, the excretion disk will
be located closer to the central star, and the ionization front would be able to penetrate to a greater extent into the disk, 
since the ionizing flux  (proportional to $r^{-2}$) is much larger. Thus, the amount of photoevaporated material
is correspondingly much greater, and the lobes form closer to the central star. Here, the pressure that opposes the fast winds is higher, leading to brighter and denser lobes. Hence, similar to the external lobes, the larger the mass-loss rate, the smaller the inner lobes.

\section{Studies on the initial surface temperature}

State of the art common envelope calculations are unable to predict the properties of the remnant central star and, hence, 
the residual envelope mass remaining in the remnant and its initial surface temperature are uncertain. 
Theoretical studies suggest that the envelope of the remnant star must be below about 0.01 \Mo in order to form a PN, for otherwise, the star
will continue its evolution as a AGB star on the Hayashi track. Thus, the efficiency of the envelope ejection plays a major role in PN formation. 

Since the initial surface temperature of the remnant core is unknown, there are several options. For example, a small ($\sim$ 6,000 K) initial temperature could be chosen, but then the evolution will require a longer computation time until the PN forms as this is governed by the stellar evolution of the post-AGB star. An alternative option is to adopt an initial temperature above $\sim$ 29,000 K for which the fast wind velocity is larger that the escape velocity of the central source as we have assumed.

To determine the differences in evolution associated with the choice of initial surface temperature of the remnant core, two nearly identical cases were computed with the only difference reflecting the fact that in the first model (A3) the initial effective temperature is 12,000 K, while in the second model (A4) the temperature is 55,000 K.  Such a choice for models A3 and A4 correspond to an initial wind velocity of 173 and 1680 \kms respectively.  It can be seen in Fig. 7 that the wind in the case of model A3 is unable to expand sufficiently in such a dense medium at 1000 days of the evolution, while model A4 already has formed a young PN. At 100 years, the external 
lobes of model A4 are larger than the lobes of model A3 by a factor of 2, and the inner lobes of model A3 have not yet formed as in the case of model A4. The ionizing front in model A3 forms a cone of $\sim$ 80 degrees, while in model A4 the ionizing front lies along the surface of the excretion disk. Note that the lobes of model A3 are formed by the expansion of the photo-ionized gas 
(champagne flow) rather than by the fast wind itself. That is, the PN formed by model A3 has not yet shown any evidence of hot-shocked gas.

These results, taken as an aggregate, show that a low initial surface temperature produces an important temporal bias on the nebular evolution.

\section{Comparison with Planetary Nebulae} 

In this section, the outcomes from some of our models are compared with the observations of a number of PNe. To facilitate the comparison, the emission measurement of the ionized gas for some snapshots are computed for models A1, A2, A4 and A5, and displayed in Fig. 8  and 9 together with a plot of the ionized density. The black equatorial zones represent the neutral gas. The effects of dust are not taken into account in the plots. Note that we have not attempted to model any particular nebula, but instead, have extracted some details that are important to the dynamics of PNe that have not been found in previous studies.

Very recently an inner hourglass nebula has been discovered in M2-9 (Castro-Carrizo et al. 2017; first seen in Clyne et al. 2015). This nebula is seen very close to the diffraction cross of the central star. Apparently, previous works on PN hydrodynamics did not resolve such a structure, and only some
indication is found in retrospect in Garc\'{\i}a-Segura et al. (1999) (model D, Figure 1).
Our model A1 (Figure 8  and 9, top) provides an insight of the mechanism responsible for such an inner nebulae, namely, it is the gas that is ablated and photo evaporated from the excretion disk. 
It is the same
mechanism described in the previous section that leads to the formation of the inner lobes. However, the material can be  ablated hydrodynamically without the need of photo-ionization, such as in the case of model D (Garc\'{\i}a-Segura et al. 1999). The large blobs of gas seen in the tips of the lobes in model A1 are also similar to those observed 
in M2-9. These long lived blobs also originate from  material that is being ripped from the equatorial, excretion disk, 
and piled-up at the symmetry axis. The fact that the actual binary in M2-9 has a period of 92 years (Corradi et al. 2011) is difficult to reconcile with a common envelope evolution unless the progenitor of M2-9 was a triple system.

IPHASXJ211420.0+434136 (also named Ou5) is a newly discovered PN which shows several nested bipolar lobes.  The light curve of the central star reveals that it is an eclipsing binary system with an orbital period of 8.74 hours (Corradi et al. 2014). The fact that it is an eclipsing binary is very important, since there 
is little confusion of projection effects. Figure 2 in Corradi et al. (2014) labels the outer and the inner lobes, as well as the cones.  
The similarities of Ou5 with our model A2 (Figure 8  and 9) are remarkable, not only in the appearance of the lobes, but also in the cones, which are computed for the first time in the literature. 
Note that our model is the result of a single, common envelope evolution event, which is able to reproduce most of the main features observed in Ou5. The existence of nested lobes in our model is due to the photoionization effect, rather than  
two episodes of mass ejection. The cones, themselves result from the gas that is photoionized close to the
excretion disk. We note that Ou5 is a high excitation nebula (Corradi et al. 2014), and the ionizing flux must be sufficiently large to ionize most (if not all) of the excretion disk.

The Skimo nebula (NGC 2392) is mainly composed of  an outer fragmented ring surrounding 
an inner nebula (Garc\'{\i}a-D\'{\i}az et al. 2012). This nebulae is seen pole on.
When a 3-D reconstructed model for the Skimo nebula 
is rotated with respect to the plane of the sky (see Figure 8 in Garc\'{\i}a-D\'{\i}az et al. 2012), the nebula resembles 
NGC 6543, the Cat's Eye nebula. Our model A4 (Figure 8) is probably a good example of an inner elliptical nebula  surrounded by an bright, fragmented disk. Note that in our case the fragmentation of the disk in the third dimension $\phi$ cannot be reproduced, but previous calculations (Garc\'{\i}a-Segura et al. 2006) showed that the ionization front is
very effective in the fragmentation process. The Cat's Eye nebula, on the other hand, is tilted
with respect to the plane of the sky, which presents two external  bipolar lobes,
a fragmented equatorial ring, and an inner elliptical nebula (Balick 2004). This morphology can be compared
with model A5 in Figure 8, where those features can be basically reproduced. Furthermore, the inner shell
is filled with hot shocked gas ($\sim 10^6$ K) produced once the wind crosses the reverse shock. This shock is visible close to the star in
our model A5, with a spherical shape. This is a very important result, since X-rays are detected inside of the inner shell of the Cat's Eye nebula (Chu et al. 2001) in very good agreement.

\section{Discussion}

The numerical results from our two dimensional hydrodynamic simulations, based on simple treatments and approximations,  reveal a rich diversity of PN structures that can emerge from a common envelope event. 
Among results is the recognition that a pole/equator density gradient is necessary prior to the common envelope event in order to produce the bipolar morphology of the outer lobes, independent of the envelope ejecta which is equatorially confined. We attribute this density gradient to the gravitational focusing effect by a companion star on the slow wind of the more evolved asymptotic giant star. The greater is the mass of the companion relative to the giant (i.e., higher mass ratio), the greater is this effect.  Such a result could be used as a basis for developing a diagnostic to constrain the nature of the binary prior to the onset of the common envelope phase. We note, however, that bipolar lobes can also be produced, in principle, by the action of jet inflated bubbles emanating from the accretion disk surrounding the companion star (Akashi \& Soker 2016). Support for the latter interpretation could be provided if indications of a jet exist (e.g., point-symmetric structures).

Many PNe show Low Ionization Structures (LIS) in their equatorial regions. We have shown, in accordance with Miszalski et al. (2009), that the formation of equatorial LIS, in the form of rings, fragmented rings, or comet-like
features is a direct result from a common envelope phase of evolution, and that they are produced by the photoionization 
of the excretion disk. In the young phases, the LIS must have had an unfragmented disk like structure, while in the later phases the fragmentation of the disk forms cometary-like shapes, as observed in the Skimo nebula. Note that the expansion velocity of the equatorial ring or LIS is proportional to the escape velocity of the progenitor star. Thus, the kinematics of these LIS can provide some constraints on (i) the radius of the giant star prior to the common envelope evolution event, (ii) separation of the binary, and (iii) the evolutionary status of the progenitor star. 
Examples of excretion discs in young phases are clearly observed in He 3-1475 and the Egg Nebula (CRL 2688).

Upon comparison of the statistics of bipolar nebulae with the distribution of elliptical PNe on the Galactic plane (Corradi \& Schwarz 1995), it is found that bipolar nebulae are larger with respect to ellipticals and characterized by higher expansion velocities. However, when individual nebulae are examined in detail 
it can be observed that there are some bipolars, which are  smaller than ellipticals, and there is no evidence for a clear correlation. As shown in \S~6, this result could be explained when the evolutionary state or the synchronization with the thermal pulses is taken into account at the moment of the common envelope event. Although the scale height of the PN above the Galactic plane is likely related to the mass of the progenitor (Corradi \& Schwarz 1995), the size of each individual nebula is not necessarily correlated with 
the scale height for post-common envelope nebulae.

Although the circumstellar nebula around $\eta$ Carinae has a similar size to that of a PN, it is by far much greater in mass. However, we point out that an outcome from our numerical results may have relevance for this nebula and others. Specifically, $\eta$ Carinae has two bipolar lobes, separated in the middle by an expanding disk (Davidson \&  Humphreys 1997). According to Smith (2017), the great eruption and the formation of the nebula could be the result of a radiation instability operating near the Eddington limit (Langer et al. 1999) or from an explosive eruption. If the latter was the case, a sudden eruptive ejection likely behaves in a similar fashion to the ejection produced in a common envelope evolution since both are catastrophic events.  In general, a large amount of mass and angular momentum are transported outwards. Accordingly, it is interesting to note the similarity of  frame B in Figure 2 with $\eta$ Carinae. At this point in time of the model evolution, the dynamics is radiative and in the so-called momentum conserving phase, similar to the case for the models in Langer et al. (1999). Thus, the formation of the excretion disk in $\eta$ Carinae could be a natural consequence of the conservation of angular momentum, and not necessarily related to the progenitor reaching a balance between the radiative acceleration, centrifugal forces, and gravity at its equator (Langer et al. 1999).

\section{Conclusions}

The structure and evolution of planetary nebulae emerging from the ejection of an envelope in the common envelope phase of binary evolution has been investigated based on two dimensional hydrodynamical simulations. The results from a suite of calculations have been presented for a range of parameters describing the temperature of the remnant core, mass loss rate prior to the common envelope event, and the mass ratio of the binary system.  From a description of the evolutions using a simplified treatment of the radiative processes and hydrodynamics it is demonstrated that the planetary nebula morphological type is dependent not only on the density structure of the ejected common envelope, but also on the degree of departure from spherical symmetry in the mass lost via a stellar wind during the AGB phase.  We have shown that the hydrodynamics associated with the photoevaporation of the matter from the excretion disk (that formed in the circumstellar environment) is especially important for the formation of the nebula shape in the innermost region. 

During the early phase of evolution (e.g., young nebulae), bipolar shapes are a natural consequence of the geometrical distribution of the common envelope ejecta. We have noted, however, that this shape is not permanent, but makes a transition to a shape characterized as either elliptical or barrel-type for older nebulae. The diversity of solutions resulting from our simulations are exemplified by cases in which nested lobes are found in which the inner region can be lobes described as bipolar, elliptical or barrel-type surrounded by bipolar outer lobes. Here, the description of the hydrodynamics of the photo-evaporated gas from the equatorial region is critical in the shape formation. In contrast to previous suggestions, we find that two separate mass ejection events are not required for the formation of a morphology described by both inner and outer lobes.  Furthermore, it is generally found that the majority of the ejected matter is concentrated in the orbital plane of the binary system and can be either in a neutral state for the early nebula phase or photoionized in the late evolutionary phase. 

In the future, higher resolution studies of the innermost region of the nebula where the circumbinary disk forms close to the central stars are planned since an important limitation to our study is that spatial resolution in the inner part of the domain is decreased as the computational grid expands.  This extra material could, in principle, play an important role in the formation of narrow waist bipolar nebulae and could be an important source of wind mass once this structure is fully evaporated.
Additionally, we also plan to consider magnetohydrodynamic effects operating on the ejected envelope and winds from the remnant core, which can facilitate the collimated structures seen, for example, in Hen 3-401 and Hen 3-1475.

\acknowledgments

We thank Michael L. Norman and the Laboratory for Computational
Astrophysics for the use of ZEUS-3D. The computations
were performed at the Instituto de Astronom\'{\i}a-UNAM at  Ensenada.
G.G.-S. is partially supported by CONACyT grant 178253.
Partial support
for this work has been provided by NSF through grants
AST-0200876, AST-0703950 and AST-1413367. FLASH Computations were carried
out using NSF Teragrid resources at the National Center for 
Supercomputing Applications (NCSA) and the Texas Advanced
Computing Center (TACC) under allocations TG-AST040024
and TG-AST040034N. P.M.R. acknowledges the Kavli Institute 
for Theoretical Physics, where some of this work was performed 
with funding by NSF under grant PHY05-51164 (the
report number for this paper is NSF-KITP-11-085). FLASH
was developed and is maintained largely by the DOE-supported
Flash Center for Computational Science at the University of
Chicago.

 \software{ZEUS-3D (version 3.4; Stone \& Norman 1992, Clarke 1996)}
 \software{FLASH (2.4; Fryxell et al. 2000)}



\clearpage

\begin{table}
\begin{center}
\caption{Model parameters. $\Mdot$ refers to the RG or AGB phase. B\&C means
Bjorkman \& Casinnelli (1993) equatorial density enhancement.}
\begin{tabular}{lccccccr}
\tableline\tableline
Model &  Type  & $M_{\rm ZAMS}$ (\Mo) & $M_{\rm core}$ (\Mo) & $\Mdot$ (\Moy) & $T_{\rm eff}$ (K) & B\&C & Figure \\
\tableline
R1 & RG & 1.05 &  0.360 & $10^{-9}$  & 29,000  &  no   & 5  \\
R2 & RG & 1.05 &  0.360 & $10^{-9}$  & 29,000  &  yes  & 2,5   \\
A1 & AGB & 1.   &  0.569 & $10^{-8}$  & 29,000  &  yes  & 3,8   \\
A2 & AGB & 2.5  &  0.677 & $10^{-8}$  & 29,000  &  yes  & 4,8   \\
A3 & TP-AGB & 1.&  0.569 & $10^{-6}$  & 12,000  &  yes  & 7   \\
A4 & TP-AGB & 1.&  0.569 & $10^{-6}$  & 55,000  &  yes  & 6,7,8 \\
A5 & AGB & 1.   &  0.569 & $10^{-7}$  & 55,000  &  yes  & 6,8   \\
A6 & AGB & 1.   &  0.569 & $10^{-8}$  & 55,000  &  yes  & 6   \\

\tableline
\end{tabular}
\end{center}
\end{table}

\clearpage

\begin{figure}
\epsscale{1.25}
\plotone{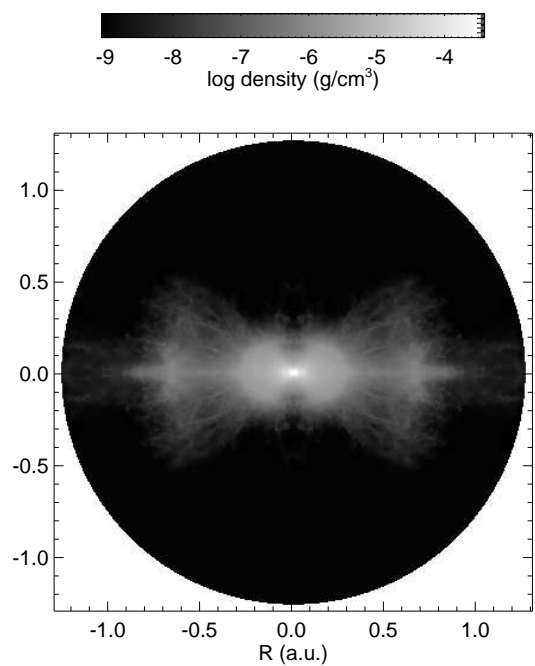}
\caption{Gas density of  the ejected envelope (initial conditions).}
\label{initial}
\end{figure}

\clearpage

\begin{figure}
\epsscale{1.40}
\plotone{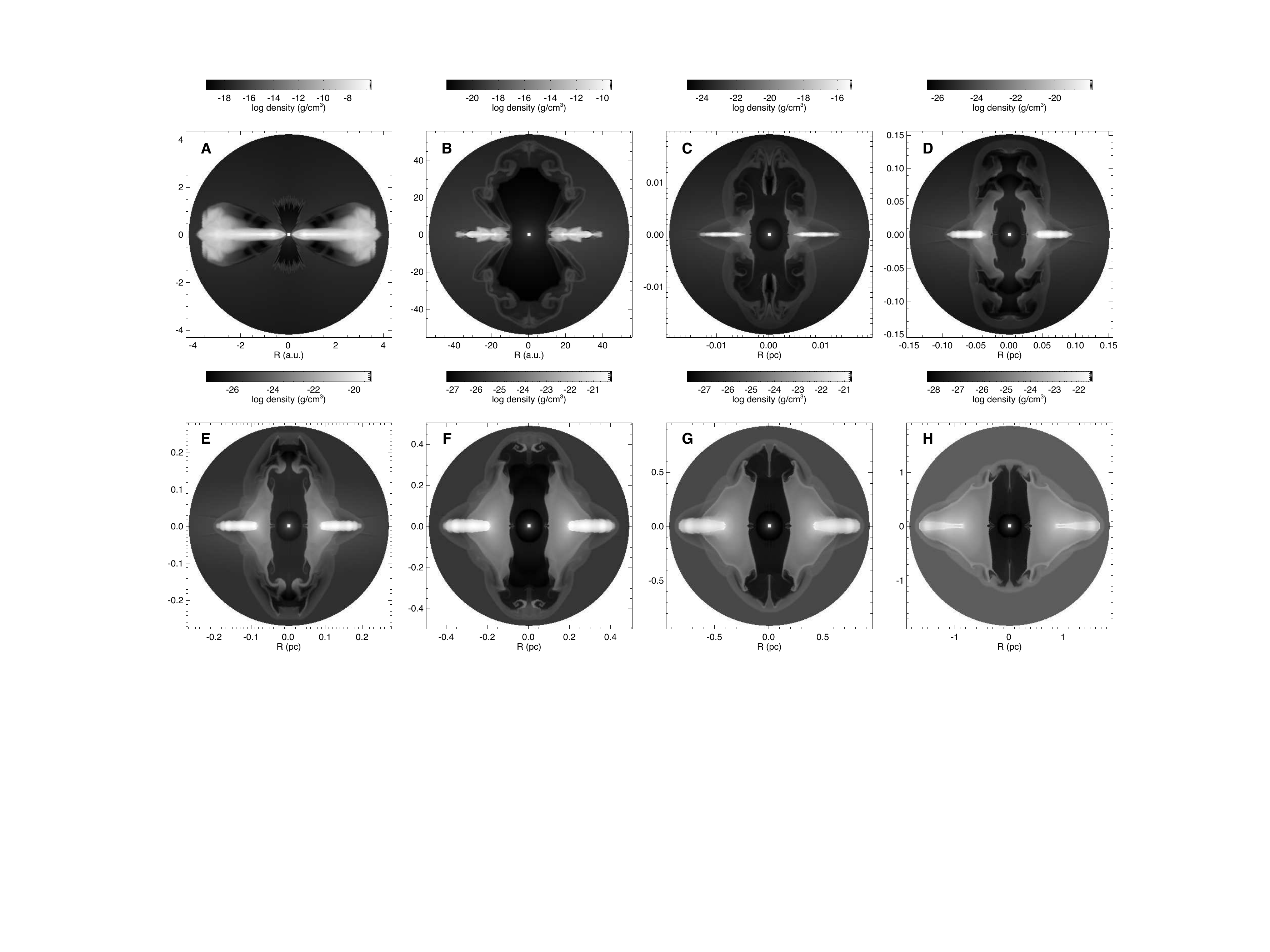}
\caption{Gas density snapshots of models R2 (RG 1.05 \Mo). Times  are (A to H): 140 , 1000  days, 100, 600, 1200, 2500, 5000 and 10,000 years. Note 
the continuous change in linear scale and greyscale. }
\label{1RG}
\end{figure}

\clearpage

\begin{figure}
\epsscale{1.40}
\plotone{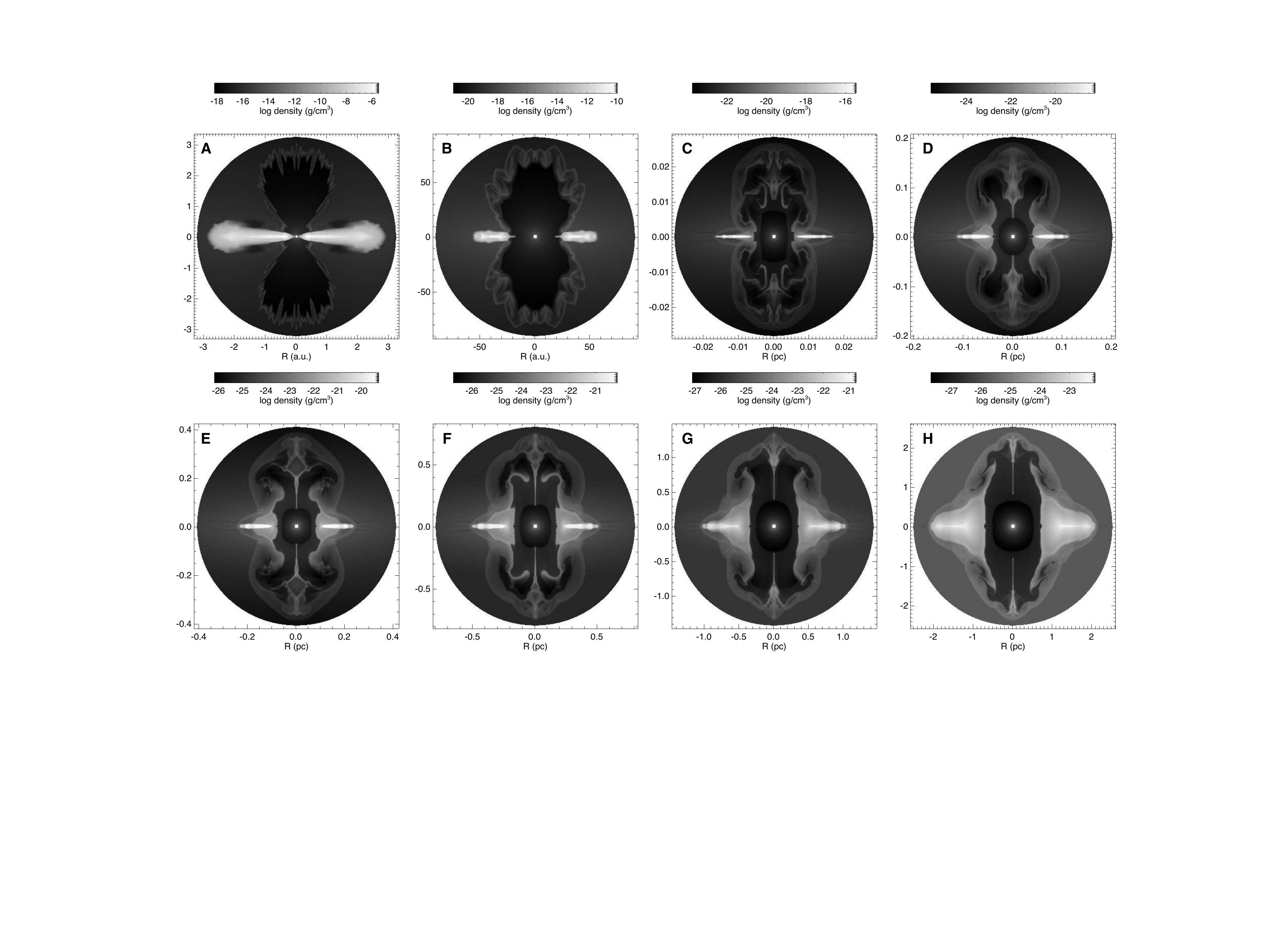}
\caption{Gas density snapshots of models A1 (AGB 1 \Mo). Times  are (A to H): 140, 1000 days, 100, 600, 1200, 2500, 5000 and 10,000 years. 
Note the continuous change in linear scale and greyscale.}
\label{1AGB}
\end{figure}

\clearpage

\begin{figure}
\epsscale{1.40}
\plotone{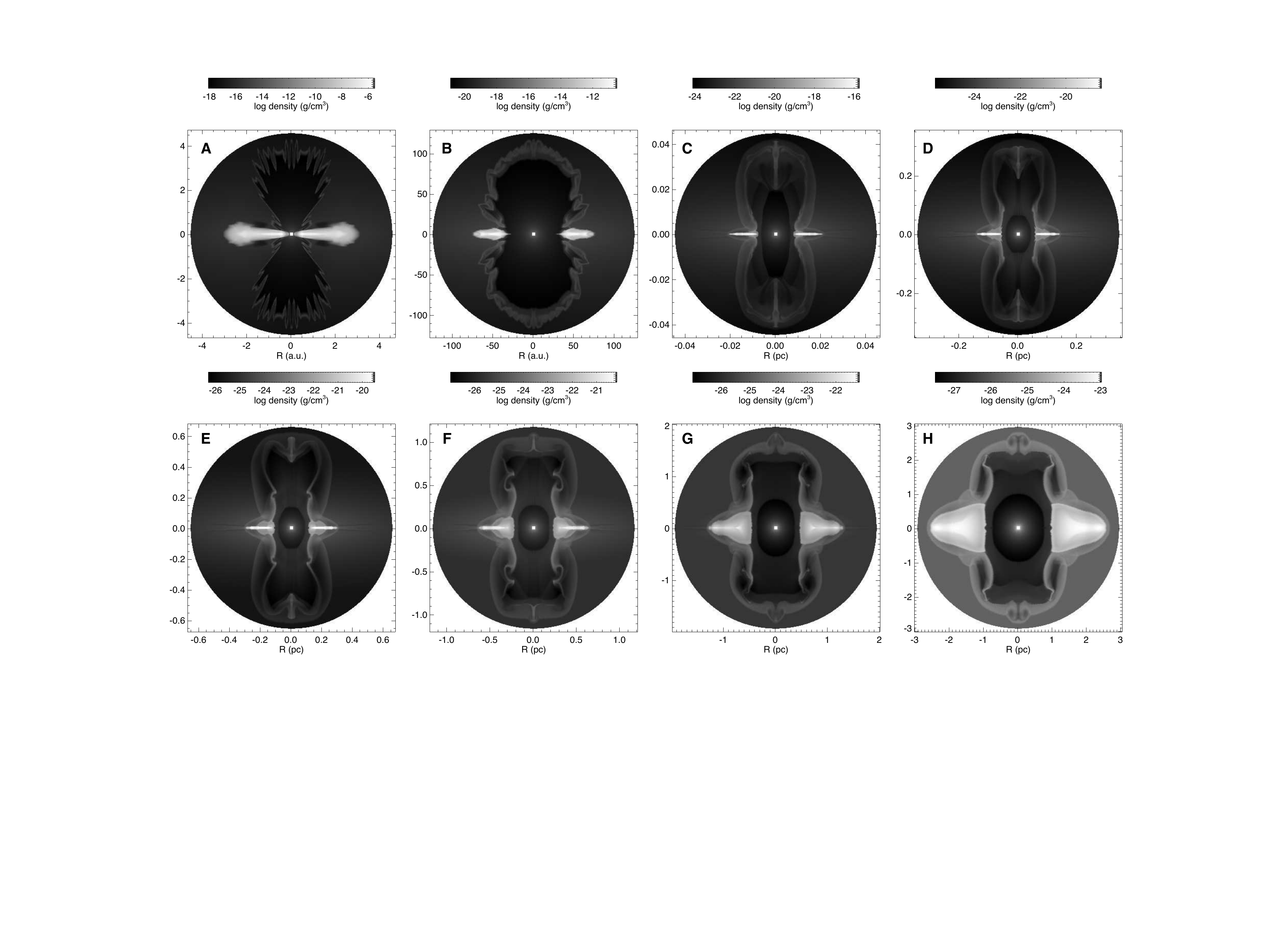}
\caption{Gas density snapshots of models A2 (AGB 2.5 \Mo). Times  are (A to H): 140, 1000 days, 100, 600, 1200, 2500, 5000 and 10,000 years. Note
the continuous change in linear scale and greyscale.}
\label{2AGB}
\end{figure}

\clearpage

\begin{figure}
\epsscale{1.25}
\plotone{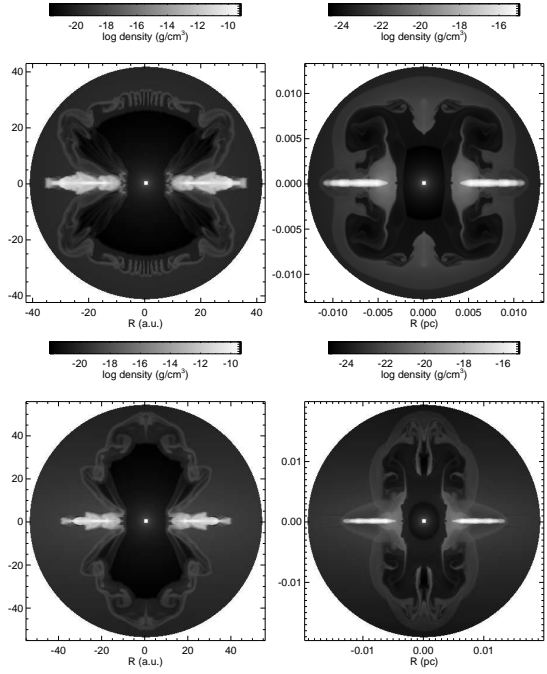}
\caption{Gas density snapshots of models R1 (top) with a small mass ratio and R2 (bottom) with a large mass ratio. Times  are (left to right): 
 1000 days and 100  years. The effect on the mass ratio is notable. }
\label{massratio}
\end{figure}

\clearpage

\begin{figure}
\epsscale{1.25}
\plotone{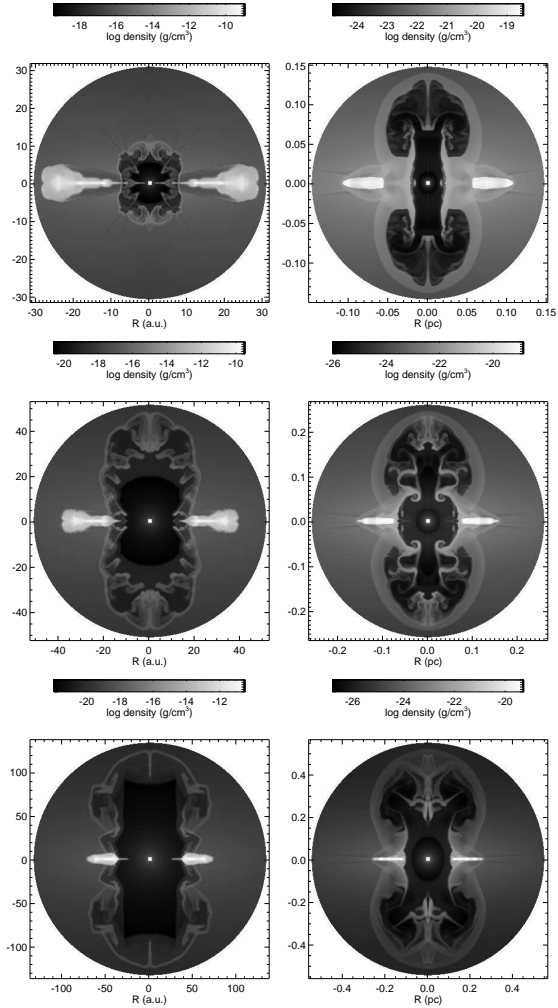}
\caption{Gas density snapshots of models A4 ($\Mdot_{\rm AGB} = 10^{-6}$ \Moy) (top), A5 ($\Mdot_{\rm AGB} = 10^{-7}$
\Moy) (center) and A6 ($\Mdot_{\rm AGB} = 10^{-8}$ \Moy)  (bottom).  Times  are (left to right):  1000 days and 1000 years. }
\label{Mdot}
\end{figure}

\clearpage

\begin{figure}
\epsscale{1.25}
\plotone{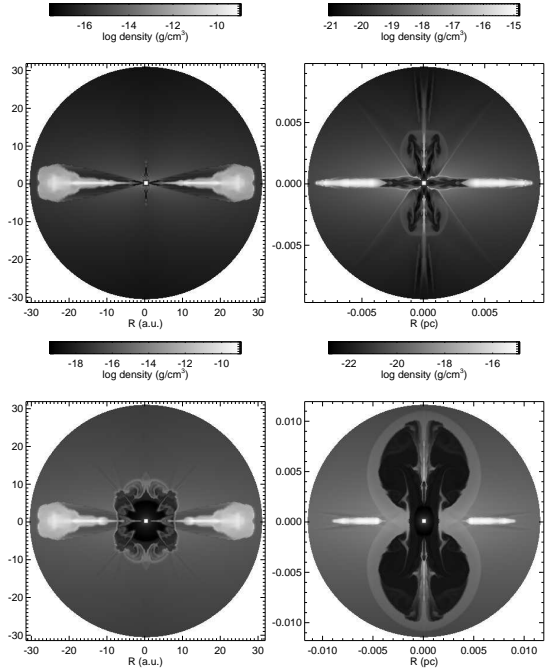}
\caption{Gas density snapshots of models A3 ( Teff = 12,000 K) (top) and A4 ( Teff = 55,000 K)  (bottom).  
Times  are (left to right):  1000 days and 100 years. }
\label{Teff}
\end{figure}

\clearpage

\begin{figure}
\epsscale{1.25}
\plotone{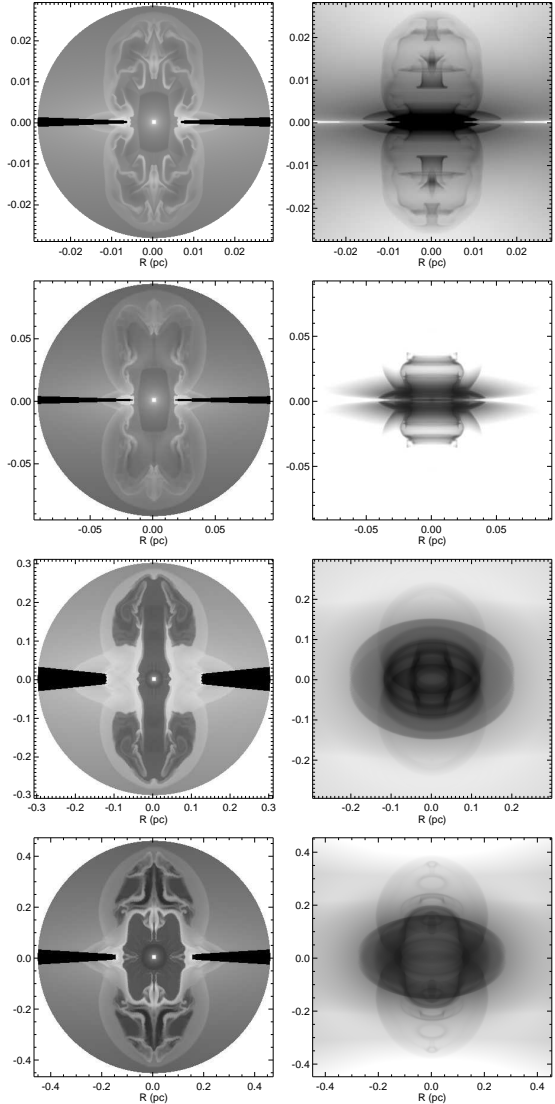}
\caption{Photoionized gas density (left) and emission measurement (right) of four 
snapshots from models A1, A2,  A4, and A5 (top to bottom).     
The first case shows inner structures similar to M2-9.
The similarities with the PN IPHASXJ211420.0+434136 are remarkable in the second case.
The other two examples, tilted 40 degrees,  show  elliptical shapes inside  bipolars, similar to the Cat's eye nebula. Right panels are reversed in grey scale. }
\label{PNmodels}
\end{figure}

\clearpage

\begin{figure}
\epsscale{1.25}
\plotone{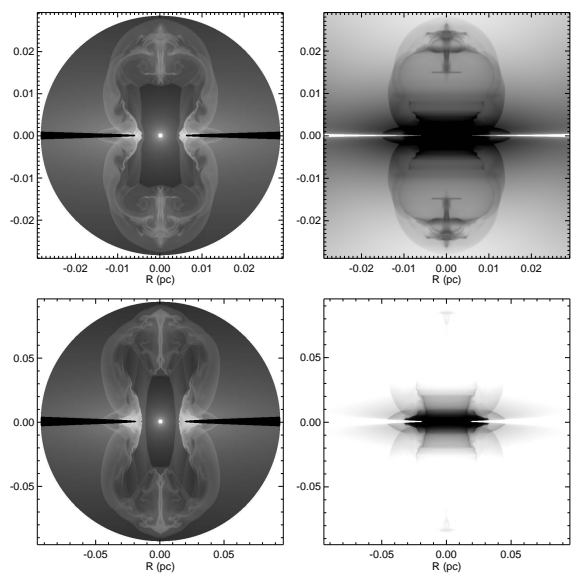}
\caption{ Photoionized gas density (left) and emission measurement (right) of two 
snapshots from models A1, A2 (top to bottom) with twice the 
resolution than Figure 8.}
\label{PNmodels}
\end{figure}


\begin{references}
     
\reference{} Akashi, M., \& Soker, N. 2016, \mnras, 462, 206
\reference{} Balick, B. 1987, \aj, 94, 671
\reference{} Balick, B. 2004, \aj, 127, 2262	
	NGC 6543. I. Understanding the Anatomy of the Cat's Eye
\reference{} Bjorkman, J. E.,\& Cassinelli, J. P. 1993, \apj, 409, 429 
\reference{} Bermudez et al. , in preparation
\reference{} Calvet, N., \& Peimbert, M. 1983, RMA\&A, 5, 319 
\reference{} Castro-Carrizo, A., Bujarrabal, V., Neri, R., Alcolea, J., S\'anchez Contreras, C., 
             Santander-García, M., \& Nyman, L.-A. 2017, A\&A, 600, 4     
\reference{} Chu, Y.-H., Guerrero, M. A., Gruendl, R. A., Williams, R. M., \& Kaler, J. B. 2001, \apj, 553, 69
\reference{} Clarke, D. A. 1996, \apj, 457, 291
\reference{} Clyne, N., Akras, S., Steffen, W., Redman, M. P., 	 Gon\c calves, D. R., \& Harvey, E. 2015, 
\reference{} Corradi, R. L. M., Balick, B., \&  Santander-García, M. 2011, A\&A, 529, 43     
\reference{} Corradi, R. L. M., Rodr\'{\i}guez-Gil, P., Jones, D., Garc\'{\i}a-Rojas, J., Mampaso, A.,
Garc\'{\i}a-Alvarez, D., Pursimo, T., Eenm\"ae, T., Liimets, T., \& Miszalski, B.	2014, \mnras, 441, 2799
\reference{} Corradi, R. L. M., Schwarz, H. E. 1995, A\&A, 293, 871
\reference{} Dalgarno, A. \& McCray, R. A. 1972, ARAA, 10, 375    
\reference{} Davidson, K., \& Humphreys, R. M. 1997, ARA\&A, 35, 1
\reference{} De Marco, O. 2009, \pasp, 121, 316
\reference{} De Marco, O., Passy, J.-C., Frew, D. J., Moe, M., \& Jacoby, G. H. 2013, \mnras, 428, 2118
\reference{} Frank, A. \& Mellema, G. 1994, \aap, 289, 937
\reference{} Fryxell, B., Olson, K., Ricker,  P., et al. 2000, \apjs, 131, 273     
\reference{} Garc\'{\i}a-D\'{\i}az, Ma. T., L\'opez, J. A., Steffen, W., Richer, M. G. 2012, \apj, 761, 172
\reference{} Garc\'{\i}a-Segura, G. 1997, \apjl, 489, 189
\reference{} Garc\'{\i}a-Segura, G., Langer, N., R\'o\.zyczka, M., \& Franco, J. 1999, \apj, 517, 767
\reference{} Garc\'{\i}a-Segura, G., L\'opez, J. A., \& Franco, J. 2005, \apj, 618, 919  
\reference{} Garc\'{\i}a-Segura, G., L\'pez, J. A., Steffen, W., Meaburn, J., \& Manchado, A. 2006, \apj, 646, 61
\reference{} Garc\'{\i}a-Segura, G., Villaver, E.,  Langer, N., Yoon, S.-C., \& Manchado, A. 2014, \apj, 783, 74
\reference{} Garc\'{\i}a-Segura, G., Villaver, E., Manchado, A., Langer, N., \& Yoon, S.-C. 2016, \apj, 823, 142
\reference{} Gawryszczak, A. J., Mikolajewska, J., \& R\'o\.zyczka, M. 2002, \aap, 385, 205
\reference{} Huckvale, L., Prouse, B., Jones, D., Lloyd, M., Pollacco, D., L\'opez, J. A., O'Brien, T.J.,      Sabin, L., \& Vaytet, N. M. H. 2013, \mnras, 434, 1505
\reference{} Icke, V. 1988, \aap, 202, 177
\reference{} Icke, V., Preston, H. L. \& Balick, B. 1989, \aj, 97, 462
\reference{} Ignace, R., Cassinelli, J. P., \& Bjorkman, J. E. 1996, \apj, 459, 671
\reference{} Kwok, S., Purton, C. R., \& Fitzgerald, P. M. 1978, \apjl, 219, 125     
\reference{} Langer, N., Garc\'{\i}a-Segura, G.,  \& Mac Low, M.-M. 1999, \apjl, 520, 49
\reference{} Livio, M. 1993, in IAU Symp. No 155 ``Planetary Nebulae'', ed. R. Weinberger and A. Acker, (Kluwer Academic Publishers, Dordrecht), 279
\reference{} Livio, M. 1994, in Circumstellar Media in Late Stages of Stellar Evolution, ed. R Clegg, P. Meikle, \& I. Stevens (Cambridge: Cambridge                Uni. Press), 35
\reference{} MacDonald, J. \& Bailey, M. E. 1981, MNRAS, 197, 995
\reference{} Matt, S., Balick, B., Winglee, R., \& Goodson, A. 2000, \apj, 545, 965
\reference{} Mellema, G., Eulderink, F. \& Icke, V. 1991, \aap, 252, 718  
\reference{} Miszalski, B., Acker, A., Parker, Q. A., \& Moffat, A. F. J. 2009, A\&A, 505, 249
\reference{} Mitchell, D. L., Pollacco, D.,  O'Brien, T. J., Bryce, M.,
            L\'opez, J. A., Meaburn, J., \& Vaytet, N. M. H. 2007, \mnras, 374, 1404    
\reference{} Mohamed ,S., \& Podsiadlowski, Ph. 2012, Baltic Astronomy, 21, 88
\reference{} Nordhaus, J., \& Blackman, E.~G.\ 2006, \mnras, 370, 2004
\reference{} Podsiadlowski, Ph., \& Mohamed, S. 2007, Baltic Astronomy, 16, 26 
\reference{} Raymond, J. C. \& Smith, B. W. 1977, ApJS, 35, 419
\reference{} Reimers, D. 1975, Mem. Soc. Liege, 8, 369
\reference{} Ricker, P. M., \& Taam, R. E. 2012, \apj, 746, 74
\reference{} Sandquist, E. L., Taam, R. E., Chen, X., Bodenheimer, P., \& Burkert, A.  1998, \apj, 500, 909
\reference{} Shu, F. M., Anderson, L., \& Lubow, S. M. 1979, \apj , 229, 223
\reference{} Smith, N. 2017, \mnras, 471, 4465
\reference{} Soker, N. 1997, \apjs, 112, 487
\reference{} Soker, N. 1998, \apj, 496, 833
\reference{} Stone, J. M. \& Norman, M. L. 1992, \apjs, 80, 753
\reference{} Taam, R. E., \& Sandquist, E. L. 2000, \araa, 38, 113
\reference{} Villaver, E., Garc\'{\i}a-Segura, G., \& Manchado, A. 2002a, \apj, 571, 880
\reference{} Villaver, E., Manchado, A., \& Garc\'{\i}a-Segura, G. 2002b, \apj, 581, 1204
\reference{} Villaver, E., Manchado, A., \& Garc\'{\i}a-Segura, G. 2012, \apj, 748, 94
\reference{} Vassiliadis, E., \& Wood, P. 1993, \apj, 413, 641
\reference{} Vassiliadis, E., \& Wood, P. 1994, \apjs, 92, 125 
\reference{} Vishniac, E. T. 1994, \apj, 428, 186



\end{references}
\end{document}